# Fabrication and extreme micromechanics of additive metal microarchitectures

Sung-Gyu Kang [a, b*], Bárbara Bellón [a], Lalith Kumar Bhaskar [a], Siyuan Zhang [a], Alexander Götz [c], Janis Wirth [c], Benjamin Apeleo Zubiri [c], Szilvia Kalácska [d, e], Manish Jain [d], Amit Sharma [d], Wabe Koelmans [f], Patrik Schürch [f], Erdmann Spiecker [c], Johann Michler [d], Jakob Schwiedrzik [d], Gerhard Dehm [a], and Rajaprakash Ramachandramoorthy [a *]

[a] Max-Planck-Institut für Eisenforschung GmbH, Max-Planck-Straße 1, 40237 Düsseldorf, Germany

[b] Current address: Department of Materials Engineering and Convergence Technology (Center for K-Metals), Gyeongsang National University, Jinju-daero 501, 52828 Jinju, Republic of Korea

[c] Department of Materials Science and Engineering, Institute of Micro- and Nanostructure Research (IMN), and Center for Nanoanalysis and Electron Microscopy (CENEM), Friedrich-Alexander-Universitat Erlangen-Nürnberg, IZNF, Cauerstraße 3, Erlangen 91058, Germany

[d] Empa, Swiss Federal Laboratories for Materials Science and Technology, Laboratory for Mechanics of Materials and Nanostructures, Feuerwerkerstrasse 39, 3602 Thun, Switzerland

[e] Current address: Mines Saint-Etienne, Univ Lyon, CNRS, UMR 5307 LGF, Centre SMS, 158 cours Fauriel, Saint-Étienne, 42023, France

[f] Exaddon AG, Sägereistrasse 25, Glattbrugg 8152, Switzerland

*Corresponding author: Dr. Sung-Gyu Kang and Dr. Rajaprakash Ramachandramoorthy
E-mail: s.kang@gnu.ac.kr and r.ram@mpie.de

# Abstract

The mechanical performance of metallic metamaterials with 3-dimensional solid frames is typically a combination of the geometrical effect ("architecture") and the characteristic size effects of the base material ("microstructure"). In this study, for the first time, the temperature- and rate-dependent mechanical response of copper microlattices has been investigated. The microlattices were fabricated via a localized electrodeposition in liquid (LEL) process which enables high-precision additive manufacturing of metal at the micro-scale. The metal microlattices possess a unique microstructure with micron sized grains that are rich with randomly oriented growth twins and near-ideal nodal connectivity. Importantly, copper microlattices exhibited unique temperature (-150 and 25 °C) and strain rate (0.001–100 $s^{-1}$) dependent deformation behavior during *in situ* micromechanical testing. Systematic compression tests of fully dense copper micropillars, equivalent in diameter and length to the struts of the microlattice at comparable extreme loading conditions, allow us to investigate the intrinsic deformation mechanism of copper. Combined with the post-mortem microstructural analysis, substantial shifts in deformation mechanisms depending on the temperature and strain rate were revealed. On the one hand, at room temperature (25 °C), dislocation slip based plastic deformation occurs and leads to a localized deformation of the micropillars. On the other hand, at cryogenic temperature (-150 °C), mechanical twinning occurs and leads to relatively homogeneous deformation of the micropillars. Based on the intrinsic deformation mechanisms of copper, the temperature and strain rate dependent deformation behavior of microlattices could be explained. Our findings provide valuable insights into the intricate mechanical responses of metallic microarchitectures under dynamic and extreme temperature conditions, paving the way to potential applications such as metal microelectromechanical systems (MEMS), MEMS packaging, etc.

Mechanical metamaterials are materials that can be characterized by their specialized 3-dimensional designs. The geometry and the intrinsic properties of consisting materials primarily determine the physical and mechanical properties of the metamaterials. Early studies through computational approaches and recent studies utilizing additive manufacturing yield unique geometries often inspired by nature [1], origami and kirigami [2,3], and atomic structures [4,5] such as honeycomb, auxetic structures, open truss lattices, closed plate lattices, and triply periodic minimal surfaces showing excellent mechanical strength to weight ratio and thermal dissipation. Combined with the intrinsic plasticity of metallic materials, the metallic metamaterials exhibit superior energy absorption capabilities under external loadings, positioning them as prime candidates for damage protection in applications subjected to extreme loading conditions, such as aerospace components [6,7] and biomedical implants [8–10].

The intrinsic mechanical properties change with materials' size and environmental conditions. For example, metals at the nanoscale or with ultrafine grained microstructure exhibit a size-related phenomenon of "smaller is stronger" [11]. High strain rate conditions ($\dot{\varepsilon} \geq 1000$ s$^{-1}$) typically leads to a dramatic increase in macroscale metal strength, owing to a viscous-drag effect of dislocation [12,13]. At cryogenic temperature, a decrease in stacking fault energy often leads to deformation induced nano-twins in the materials [14,15]. Thus, the mechanical response of the macroscale metals is well-known to be dependent on the size- and loading condition. However, for metallic metamaterials especially at the mesoscales and smaller, a rigorous study of their deformation behavior and mechanical properties under different loading conditions that may result from the combined influence of geometrical and intrinsic properties is largely unexplored.

The key reason for this limitation stems from the lack of fabrication methods to manufacture mesoscale 3D metallic metamaterials with micrometer resolution. Conventional powder- and energy deposition-based additive manufacturing processes typically lack resolution for fabricating mesoscale architectures with a component size scale below 50 μm. Alternatively, top-down microfabrication of planar 2.5-dimensional microarchitectures from bulk substrates and thin films of typically silicon and a few other selective materials such as nickel, tantalum [16], and aluminum nitride [17] is optimized via the widely known ultraviolet (UV) based lithographic method. In these lithographic methods, the selection of materials is limited, and planar processing significantly restricts the geometric freedom in fabricating complex mechanical components that require high-aspect ratio three-dimensional geometries. Few studies have reported additive manufacturing of 3D metallic microarchitectures with simpler geometries such as dots, pillars, and bridges at the mesoscale via direct ink writing [18], laser-induced forward transfer [19], laser-induced photoreduction [20], and electrohydrodynamic redox [21]. Recently, hydrogel infusion-based additive manufacturing (HIAM) method has been used to fabricate more complex copper and nickel metamaterials at nano- and micro-scales [22–24]. However, the HIAM method involves several complex steps involving 3D printing polymer scaffolds, metal ion infusion, calcination and reduction to produce metallic metamaterials. Similarly, two-photon lithography (TPL) is extensively used to create templates for microarchitecture, which are subsequently filled with metal through electrodeposition [25]. Copper [26] and nickel microlattices [27,28] with complex geometries fabricated by TPL and electrodeposition show smooth surfaces and good nodal connectivity. Expectedly, these microarchitectures exhibit improved yield strengths due to the mechanical size effect of materials. However, the TPL based fabrication process is also a complex two-step process and as such possesses a few intricacies. For example. during electroplating into the template, pores can be introduced owing to the trapping of bubbles in the straight struts and microlattice struts in the previous studies showed elliptical cross-sections originating from the voxel shape of the laser based TPL process. It is therefore essential to identify a simple microfabrication methodology that will allow true freeform 3D printing of complex metallic microarchitectures with controllable microstructure.

This study presents the temperature- and rate-dependent mechanical responses of microscale metal metamaterials originating from the intrinsic microstructure dependent properties of the base material. For fabrication, a pioneering approach was taken using a localized electrodeposition in liquid (LEL) process, enabling high-precision creation of full-metal copper structures with component sizes ranging from a few microns to hundreds of microns (Fig. 1(a)). LEL is a single-step fabrication technique that enables the consistent manufacture of full-metal 3D microarchitectures with complex geometries. As a model study, the structural and microstructural investigations were conducted on copper microlattice with an octet geometry. Furthermore, to explore the mechanical performance under various conditions, *in situ* mechanical tests of the copper microlattice were performed at a wide range of strain rates from 0.001 to 100 s$^{-1}$ and various temperature conditions (-150 and 25 °C). To elaborate on the unique deformation behavior and the mechanical performance of the microlattices, copper micropillars with the same size scale as the microlattices' struts were fabricated and mechanically tested under the same conditions. The pre- and post-mortem structural and microstructural analysis of copper micropillar explains that the intrinsic deformation mechanism of copper under various conditions controls the mechanical response of metamaterials. The results especially highlight the superior yet tunable mechanical properties of copper microlattices under extreme loading conditions, demonstrating their potential as structural parts for dynamic microscale applications.

**Copper microlattice fabrication and microstructure**

We chose the octet truss design, which is known for its low mass density and good structural rigidity originating from its high nodal connectivity, as the unit-cell geometry. The unit-cell length and the strut diameter were chosen as 15 µm and 2.5 µm, respectively. The microlattice was constructed by repeating the unit cell three times along each coordinate axis (Fig. 1(b)). To ensure uniform deformation along the loading direction, the 5-µm-thick plates were added at the bottom and top of the lattice structure. Fig. 1(c) shows the location of each voxel for printing. The diameter of the voxels and the overlap between them were set to 2.5 µm and 0.5 µm, respectively. Fig. 1(d) shows an SEM image of an as-fabricated copper microlattice with straight struts and a smooth surface, which is consistent with the original computer-aided design (CAD) model. It can be attributed to the optimized printing parameters identified for the localized electrodeposition process and the use of additives, such as levelers and grain refiners, in the electrolyte (see the Methods section). The electrodeposition is done in a voxel-by-voxel manner where the voxel location is controlled by programmable linear motors with ~10 nm precision, which enables voxelized printing of copper droplets at precise locations. Further, the printing parameters for the microlattice in this study were optimized. It was observed that the nodal connectivity and strut diameter are highly dependent on the printing parameters such as the potential for the printing chamber and pressure for ion ink flow (Fig. S1). Through trial and error, the optimal potential (-0.5 V) and pressure (25 mbar) were found. Lastly, the additives in the electrochemical ink lead to a smooth and consistent surface of the microlattice and consistent microstructure throughout the architecture. The dimples at the top plate possibly originated from the irregularities in electrodeposition. To examine the morphologic integrity and connectivity of the complete lattice structure in 3D, we conducted nano X-ray computed tomography (nano-CT). Fig. 1(e) and Mov. S1-3 display a volume rendering of the 3D reconstruction of the microlattice. As expected, SEM images of the microlattice and the representative unit cell (cubic with a red dashed line in Fig. 1(e)) in the nano-CT reconstruction from outside is consistent with the original CAD design in Fig. 1(b). X-ray absorption contrast images and slices through the nano-CT reconstruction of the microlattice in Fig. S2 also demonstrate the well-defined geometry of the microlattice. Notably, there are few struts with poor nodal connectivity at the outer surface of the microlattice (indicated with red arrows in Fig. 1(e)). This can be attributed to the printing artifacts in the LEL process. The nano-CT results demonstrate the micron-scale resolution and geometric freedom of the LEL process. Fig. 1(f) shows the diameters of arbitrarily chosen 70 struts (measured at the center) from cross-sectional nano-CT slices in Fig. S2. The average strut diameter is 2.52 ± 0.11 μm, and the relative density is about 0.28.

Fig. 1(g, h) shows the microstructure of the representative unit cross-section along the building direction of the microlattice obtained using SEM-transmission Kikuchi diffraction (TKD). The image contrast map in Fig. S3 suggests that the unindexed areas that can be observed at the node and corners of struts may originate from the ion-beam milling-induced damage. Both inverse pole figure maps (Fig. 1(g)) along and perpendicular to the building direction reveal that the microlattice consists of polycrystalline copper without strong crystallographic texture. In the grain boundary map (Fig. 1(h)), there is a small number of general high-angle grain boundaries (HAGBs, colored in blue) that are far from a twin relationship, sectioning the strut diameter into a few grains (average grain size of 1.04±0.24 μm). Importantly, there is a high fraction of Σ3 twin boundaries (TBs) within the grains. The color of the TB was set to gradually change from yellow to red, as the misorientation angle moves away from 60° by upto 10°. Apparently, for the TBs in the as-fabricated microlattice, the misorientation angle between neighboring grains is almost 60°. The formation of TBs during LEL may be attributed to the additives, as they increase the nucleation sites of copper deposition and reduce the interfacial energy, leading to the formation of TBs [29]. In summary, the copper microlattice in this study shows well-defined geometry and dense microstructure with a high fraction of TBs of 86%.

**Compression of microlattices**

To examine the deformation behavior and mechanical performance of the copper microlattices, *in situ* micromechanical tests were performed under various extreme loading conditions. Using a piezo-based actuator with an embedded strain gauge for measuring displacement and a typical strain gauge based load cell, the mechanical response of the microlattices was captured at both room and cryogenic temperatures at strain rates ($\dot{\varepsilon}$) of 0.001 to 1 s$^{-1}$. At room temperature, the rate-dependent properties of the microlattices were further captured at higher strain rates up to 100 s$^{-1}$ via a high-stiffness piezoelectric load cell, replacing the compliant strain gauge-based load cell. Note that this is the first systematic study on the deformation behavior of metallic microlattices under such extreme conditions with strain rates spanning 5 orders of magnitudes at room temperature (RT) and 3 at cryogenic temperatures (CT). There are only two previously reported quantitative studies on the room temperature compression of polymer microlattices at $\dot{\varepsilon}$ upto 1000 s$^{-1}$ [30,31]. Detailed information on the mechanical testing system used in the current study is given elsewhere [32]. Fig. 2(a, b) shows SEM images of the microlattices during compression at a $\dot{\varepsilon}$ of 0.01 s$^{-1}$ at both RT and CT. Regardless of temperature, the microlattice seems to exhibit homogeneous deformation overall, but the struts of the microlattice show slightly different deformation behavior depending on temperature. At RT, the deformation starts with the buckling of struts at the bottom plate (indicated by white arrows in Fig. 2(a)). It has been previously reported that the deformation of the lattice starts from the struts with low nodal connectivity, that is, the struts at the surface or corner [33]. At CT, on the other hand, there was no noticeable buckling of surface or corner struts. Fig. 2(c) shows the representative stress-strain curves of microlattices compressed at different temperatures and $\dot{\varepsilon}$ conditions (2-3 microlattices were tested at each condition). The copper microlattice shows higher yield strength (and flow stress at 0.2 engineering strain – Fig. S4) at CT than RT as plotted in Fig. 2(d). The buckling of struts at RT can be captured by a stress decrease after the elastic deformation in the stress-strain curve. The curves in the insets obtained at 0.001, 0.01, and 0.1 s$^{-1}$ clearly show that the stress decrease occurred only at RT. In addition, as shown in Fig. 2(c) (and more in detail in Fig. S5), except for the microlattice compressed at 100 s$^{-1}$, all microlattices compressed at RT exhibit the stress overshoot after the elastic deformation in the stress-strain curves, while those compressed at CT do not. This stress decrease, also known as post-yield softening, is known to occur in the lattice structure due to an abrupt buckling or fracture of struts [34–36]. At RT, 100 s$^{-1}$ compression results in oscillations in the stress-strain curve beyond yield. These oscillations can be attributed to the activation of the resonance in the piezoelectric load cell, which is in turn the result of the change in stiffness of the sample (for example at the elastic-plastic transition) during deformation at high $\dot{\varepsilon}$ [37].

A clear rate-dependency can be seen from the stress-strain signatures of the microlattices. For example, at RT, the yield strength at an offset of 0.2% (Fig. 2(d)) increases with $\dot{\varepsilon}$ until ~1 s$^{-1}$ (107.1 ± 7.3 MPa at 0.001 s$^{-1}$ and 160.8 ± 2.6 MPa at 1 s$^{-1}$). Beyond a $\dot{\varepsilon}$ of 1 s$^{-1}$, the yield strength almost saturates. At CT, on the other hand, yield strength shows negligible change with $\dot{\varepsilon}$ from 0.001 to 1 s$^{-1}$ and is almost constant at around 230 MPa. The rate sensitivity factor was extracted based on Eq. (1) below:

$$m = \frac{\partial \ln \sigma_y}{\partial \ln \dot{\varepsilon}} \tag{1}$$

where $\sigma_y$ is the yield strength. At RT, the microlattice shows $m$ value of $5.2 \times 10^{-2} \pm 0.8 \times 10^{-2}$ up to a $\dot{\varepsilon}$ of 1 s$^{-1}$ and then shows a decreased value of $4.0 \times 10^{-4} \pm 8.0 \times 10^{-4}$ up to a $\dot{\varepsilon}$ of 100 s$^{-1}$, implying negligible rate dependence at RT for $\dot{\varepsilon} \geq 1$ s$^{-1}$. At CT, the microlattice shows a low $m$ value of $1.5 \times 10^{-3} \pm 1.7 \times 10^{-3}$ up to a $\dot{\varepsilon}$ of 1 s$^{-1}$, implying again negligible rate dependence. Different $m$ values suggest that the deformation mechanism of microlattice changes with temperature and $\dot{\varepsilon}$.

Change in the deformation mechanism of microlattice can also be confirmed via the compressed shape. Fig. 2(e, f) shows the compressed SEM images of microlattices. At RT (Fig. 2(e)), the struts at the surface of the microlattice bulged out perpendicular to the loading direction at high $\dot{\varepsilon}$ ($\geq 1$ s$^{-1}$). At CT (Figure 2(f)), on the other hand, the bulging out of the surface struts occurred regardless of $\dot{\varepsilon}$. The degree of bulging out has been calculated (Fig. S6) and shows clear dependencies on $\dot{\varepsilon}$ and temperature, which mimics closely the dependencies in the yield strength (rate dependency only exists at RT and $\dot{\varepsilon} \leq 1$ s$^{-1}$). Moreover, compared to the struts before compression in Fig. 1(c), the struts of microlattices after the compression at RT show clear slip steps, as indicated by the white arrows in Fig. 2(e). The slip steps are easier to find in the microlattices compressed at rather low $\dot{\varepsilon}$, suggesting less localized deformation of the strut occurred at high $\dot{\varepsilon}$ ($> 1$ s$^{-1}$). In the case of microlattices compressed at CT, they show smoother surfaces and thicker diameters, indicating more homogeneous deformation of struts at CT than RT. The above analysis on the yield strength, the degree of bulging out, and the slip steps at the struts of compressed microlattices suggest that the deformation behavior of microlattices in this study can be categorized into three regimes: RT 0.001–1 s$^{-1}$, RT 1–100 s$^{-1}$, and CT 0.001–1 s$^{-1}$.

Given the intended future applications of these copper microlattices in impact protection, the absorbed energy of the microlattices with respect to density was calculated and compared against the absorbed energies of other microlattices previously reported in the literature (Fig. 2(g)). The absorbed energy ($U_A$) can be obtained as below:

$$U_A = \int_0^{\varepsilon_d} \sigma \, d\varepsilon \tag{2}$$

where the $\sigma$, $\varepsilon$, and $\varepsilon_d$ are the stress, strain, and densification strain, respectively. For simplicity, $U_A$ of microlattices compressed at RT 0.001, 1, 100 s$^{-1}$ and CT 0.001, 1 s$^{-1}$ are plotted in Fig. 2(g). The $U_A$ of microlattices is 1.5~2.0 times higher at CT than RT, as expected from the stress-strain curves in Fig. 2(c). Importantly, compared to the copper microlattices from a previous study [26], the microlattices in this study show comparable $U_A$ even with lower relative density. The high energy absorption per unit of mass can be attributed to the innate microstructure, the accurate dimensions of the strut, and the excellent nodal connectivity in the microlattices used in our current study. First, due to the small grain size, and a large number of TBs and grain boundaries in the microlattices of this study, high yield strength and strong hardening behavior during deformation are expected. Second, the microlattices in this study possess struts with a near-circular cross-sectional shape, which was not the case in the previous study of Gu *et al* [26]. A strut with a circular cross-section would lead to relatively even stress distribution at the strut-node intersection during deformation compared to that with an elliptical cross-section. Finally, the near-ideal nodal connectivity, which is close to the original design, would lead to high strength and high energy absorption capability of the microlattice. To investigate the effect of nodal connectivity, we conducted an additional compression test of a copper microlattice printed intentionally with poor connectivity (Fig. S7). In this case, the deformation is concentrated at the weak nodes and leads to a localized failure. The corresponding yield and flow stresses obtained from the stress-strain signatures (Fig. S7(b)) are only a quarter of those of the microlattices with good nodal connectivity. Further, to determine whether the existence of a top plate affects the deformation behavior and also $U_A$, finite element methods (FEM) based simulations were conducted. The compressive load-displacement curves of microlattices with and without the top plate show no significant difference (Fig. S8). Thus, the copper microlattice structures in this study exhibit superior mechanical properties due to their unique microstructure and precise geometry that is near-identical to the original CAD model. Similar reasoning can be used to explain the fact that the copper microlattices in this study can achieve an energy absorption per unit mass of 20 J/g, which is significantly better than the microlattices made of other metals (5 J/g for Aluminum and Nickel [38], and 3.5 J/g Gold [39]), polymer (10 J/g) [40], metal-polymer composite (2 J/g) [41,42], and for alumina (3.5 J/g) [43].

**Base material characterization – Copper micropillar compression**

To understand the deformation mechanisms of the microlattice, the intrinsic deformation mechanism of base material copper needs to be investigated. Previous research has shown that struts of octet-truss structures are subjected to compressive or tensile stresses depending on their orientation with respect to the loading direction [44–46]. Despite the non-uniform stress distribution within the struts and the presence of stress concentrations at the nodes, the overall deformation of the octet-truss structure is primarily governed by the compressive and tensile deformation behavior of struts. This implies that the deformation mechanisms of the microlattice can be understood by examining the behavior of micropillars with the same microstructure when subjected to uniaxial compression at comparable rates and temperatures. It should be noted that a FEM simulation confirmed that the global strain rate experienced by the microlattices is quite analogous to the strain rate locally in the struts and nodes (except for at the struts in the exposed outer surfaces – Fig. S9).

The copper micropillars were also fabricated by LEL. Fig. 3(a, b) shows a computer-aided design and voxel locations of the dog-bone-shaped micropillar with a diameter of 3 μm and a length of 10 μm in a gauge section, ably mimicking the dimensions of the struts used in the microlattices. Via the LEL process, arrays of micropillars were fabricated on the substrate. As-fabricated micropillars show a smooth surface and a straight gauge section which are again consistent with the original CAD model (Fig. 3(c)).

For microstructural analysis, the cross-section of the micropillar parallel to the loading direction was sampled via focused ion beam milling. Fig. 3(d) shows the inverse pole figure (IPF) maps along and perpendicular to the building direction and the grain boundary (GB) map of the gauge section of the as-built micropillar obtained using TKD. The IPF map indicates the copper micropillar is indeed polycrystalline and the GB map with HAGBs (colored in blue) shows that there are a few grains across the pillar diameter, which is consistent with the strut of the microlattices. Importantly, a high fraction of Σ3 boundaries was also present in the grains and the boundaries' misorientation angle distribution was analogous to those present in the microlattices (Fig. S10), confirming the similarity of microstructure. Via an *in situ* micromechanical testing system with the same piezo-based actuator used for testing the microlattices, compression tests were conducted at room/cryogenic temperatures and a wide range of $\dot{\varepsilon}$.

When the temperature changes, copper micropillars also show different deformation behavior. Fig. 3(e) shows the stress-strain curves of micropillars. The curves show higher yield strength and flow stress at CT than at RT. Fig. 3(f) shows the summary of 0.2% strain offset yield strength of copper micropillars at different $\dot{\varepsilon}$ and temperatures. Depending on $\dot{\varepsilon}$, the yield strength is 1.1–1.4 times higher at CT than at RT. Moreover, the yield strength at RT increases with $\dot{\varepsilon}$ when $\dot{\varepsilon} \leq 1$ s$^{-1}$. When $\dot{\varepsilon} \geq 1$ s$^{-1}$, there is no noticeable change in the stress-strain curves with $\dot{\varepsilon}$. Analogous to the microlattice, the micropillars compressed at RT and $\dot{\varepsilon} \leq 1$ s$^{-1}$ exhibit $m$ value of $3.1 \times 10^{-2} \pm 0.1 \times 10^{-2}$, while at $\dot{\varepsilon} \geq 1$ s$^{-1}$, the micropillars exhibit a significantly low $m$ value of $8.0 \times 10^{-4} \pm 3.0 \times 10^{-4}$, implying negligible strain rate dependence. At CT, the micropillars exhibit a low $m$ value of $4.8 \times 10^{-3} \pm 1.5 \times 10^{-3}$ regardless of the value of $\dot{\varepsilon}$, implying almost insignificant rate dependency. A similar trend was found in the flow stress obtained at an engineering strain of 0.2 (Fig. S11). Interestingly, there were also stress drops in the stress-strain curves of micropillars compressed at CT at $\dot{\varepsilon}$ of 0.001, 0.01, 0.1 s$^{-1}$ (Inset in Fig. 3(e)). These stress drops identified at CT will be discussed in detail in the following section.

The compressed shape (Fig. 3(g) and S12) of the micropillars also suggests changes in the deformation behavior at different rates and temperatures. At RT, when $\dot{\varepsilon} \leq 1$ s$^{-1}$ the compressed micropillars show clear slip steps on the surface, suggesting a localized deformation. When $\dot{\varepsilon} \geq 1$ s$^{-1}$, the micropillars exhibit a mixed-mode deformation, where not only slip steps appear on the surface but also partial barreling-like homogeneous deformation around the slip step. At CT, regardless of $\dot{\varepsilon}$, the compressed micropillars show relatively homogeneous deformation without any localized slip step on the surface. Further clues to the responsible deformation mechanisms were obtained using post-mortem microstructural analysis. The cross-sections of compressed micropillars parallel to the loading direction were obtained via focused-ion beam milling.

**Temperature and rate responsive deformation mechanism of copper micropillars**

Compression leads to major changes in the microstructure of the micropillars. Fig. 4(a-c) shows microstructural analysis of the compressed micropillars. The IPF map along the building direction and the GB map of the gauge section of the micropillar compressed at RT and $\dot{\varepsilon}$ of 0.001, 1, and 100 s$^{-1}$ (Fig. 4(a)) indicates that, at every $\dot{\varepsilon}$, the straight growth TBs decrease after the compression. The GB maps indicate HAGBs and defective TBs in the gauge section of the compressed micropillars. Here, the defective TB is defined as the TB where the misorientation angle between neighboring grains deviates at least by 5° from 60° but less than 10°. Fig. 4(d) shows misorientation angle distribution of the undeformed and compressed micropillars. Fractions of TBs, defective TBs, and HAGB can be extracted from distribution plot. The undeformed micropillar exhibits high TB fraction of 86%, negligible defective TB fraction of 1%, and HAGB fraction of 12%. Notably, the micropillars compressed at RT and $\dot{\varepsilon}$ of 0.001 and 1 s$^{-1}$ commonly show decreased TB fraction of 50% and 45%, respectively. Furthermore, a fraction of defective TBs increases to 11% and 14%, and a corresponding fraction of HAGBs increases to 33% and 31%, respectively.

A change in the microstructure induced by compression at RT can be attributed to the slip transmission mode of the TB. Specifically, a "hard" mode transmission, where the Burgers vector of the dislocation in the first grain can not be maintained geometrically in the second grain, can lead to a dislocation pile-up and dislocation networks changing the misorientation angle between twin related grains [47–50]. Further deformation may lead to twin broadening or shrinkage by moving partials next or in the twin plane or dislocation incorporation into the TB can cause its transformation into a HAGB [51,52], which was observed in the micropillars compressed at RT. Notably, the region with increased HAGBs and the defective TB varies with $\dot{\varepsilon}$. As can be estimated from Fig. 4(a), the area with defective TBs is 1.3 times larger in the micropillar compressed at RT and $\dot{\varepsilon}$ of 100 s$^{-1}$ than at $\dot{\varepsilon}$ of 0.001 and 1 s$^{-1}$. It suggests that at $\dot{\varepsilon} = 100$ s$^{-1}$ the deformation is relatively homogeneous with very few localizations throughout the gauge section of the micropillar, which is consistent with the SEM image in Fig. 3(g).

Microstructural change in the micropillars compressed at CT implies a deformation mechanism different from the mechanism observed at RT. Fig. 4(b) shows the IPF map along the building direction and the GB map of the gauge section of the micropillar compressed at CT and $\dot{\varepsilon} = 0.001$ and 1 s$^{-1}$. The microstructures of micropillars commonly show HAGBs and TBs. It is critical to note that there are straight defective TBs in the GB maps. In the case of the micropillar compressed at CT and $\dot{\varepsilon} = 0.001$ s$^{-1}$, these TBs are located at the center and upper part of the gauge section. In the micropillar compressed at CT and $\dot{\varepsilon} = 1$ s$^{-1}$, these TBs are mostly found in the right part of the pillar. The corresponding misorientation distribution in Fig. 4(d) indicates that the micropillars compressed at CT and $\dot{\varepsilon}$ of 0.001 and 1 s$^{-1}$ show decreased TB fraction of 66% and 59%, respectively. Note that the TB fractions are higher for the micropillars compressed at CT than at RT. Furthermore, a fraction of defective TBs increases to 7% and 12%, and a corresponding fraction of HAGBs increases to 22% and 21%, respectively.

A twin width ($w_{twin}$) in the compressed micropillars is another parameter that can indirectly explain the possible deformation mechanisms. Via a line intercept method, $w_{twin}$ was estimated from Fig. 3(d) and 4(a, b). The line intercept method only accounts for the appearance of TBs and an averaged distance between them, but it does not distinguish the twin broadening or shrinkage. The defective TBs were also counted as the TBs for $w_{twin}$ measurement. For the undeformed micropillars, $w_{twin}$ is $0.27 \pm 0.02$ μm. For the micropillars compressed at RT, $w_{twin}$ increases to $0.39 \pm 0.10$ μm and $0.50 \pm 0.12$ μm when $\dot{\varepsilon} = 0.001$ and 1 s$^{-1}$, respectively. A clear change in $w_{twin}$ can be attributed to dislocation-based plasticity and resulting interaction with pre-existing TBs of copper at RT. Further identification of TB movements arising from the slip transmission and TB interactions requires significant modifications and improvement to the experimental setup, which is beyond the scope of this study.

In the case of the micropillars compressed at CT, $w_{twin}$ are $0.26 \pm 0.07$ μm and $0.29 \pm 0.06$ μm when compressed at $\dot{\varepsilon} = 0.001$ and 1 s$^{-1}$, respectively, which are only moderately different from $w_{twin}$ of the undeformed micropillar, suggesting that the pre-existing TBs are not significantly involved in the deformation process. Importantly, scanning transmission electron microscopy (STEM) analysis revealed a fine deformation twin structure in the micropillars compressed at CT and $\dot{\varepsilon} = 0.001$ s$^{-1}$, which was not resolved by SEM-TKD. Fig. 4(c) shows the high-angle annular dark field (HAADF)-STEM images of the micropillar compressed at CT and $\dot{\varepsilon} = 0.001$ s$^{-1}$ and the Fast Fourier transform (FFT) patterns from atomically resolved areas with straight TBs (Area 1 and 2 in Fig 4c, corresponding high-resolution images are displayed in Fig. S13). We specifically focused on the gauge section with straight boundaries. The FFT patterns clearly indicate that the straight boundaries are the deformation TBs with ($\bar{1}\bar{1}1$) twin planes, and the corresponding $w_{twin}$ is about 40 nm. It suggests that mechanical twinning might have occured in the micropillar during compression at CT. Although some TBs in SEM-TKD images appear to have a plane normal along the loading direction (Fig. 4(b)), further pole figure analysis revealed that their associated {111} twin planes can be ~30° angle away from the loading direction (Fig. S14). The STEM investigation revealed mechanical twins for the CT deformation. Thus, it can interpreted that at CT, the micropillar deformation is dominated by mechanical twinning rather than dislocation plasticity based mechanisms.

Combined with the microstructural analysis, the stress drop events, which occurred only at CT in Fig. 3(e), are in this case a typical signature of a mechanical twinning based deformation mechanism. It has been previously reported that mechanical twins can occur in copper when deformed at CT leading to stress drops [53,54]. Fig. 4(e) shows stress drop magnitudes (Δσ) and drop times (Δt) from the stress-strain curves. The inset plot explains, as an example, how Δσ and Δt were evaluated from the stress-strain curve at CT and $\dot{\varepsilon}$ of 0.01 s$^{-1}$. Except for a few outliers, most data points are in the range of 23< Δσ <80 MPa and 0.075< Δt <0.35 s, suggesting that the stress drop is induced by the same event. Given the short drop time, it can be deduced that at $\dot{\varepsilon} = 1$ s$^{-1}$ the stress drop is less likely to occur because the total deformation time is relatively short (about 0.2 s). Meyers *et al.* previously suggested that a slip-to-twin transition can occur at low temperatures as the stress for dislocation slip increases with decreasing temperature [55].

The postulated deformation mechanisms of copper micropillars at various temperatures and $\dot{\varepsilon}$ are schematically shown in Fig. 4(f). At RT 0.001–1 s$^{-1}$, a dislocation glide may lead to localized deformation and the slip step at the surface [56]. Such dislocation glides can then lead to dislocation-TB interactions. As another method to investigate such deformation mechanisms, thermal activation analysis was conducted. The apparent activation volume ($V^*$), a kinetic signature of the deformation mechanism, can be obtained from Eq. (3) below:

$$V^* = \sqrt{3} k_B T \frac{\partial \ln \dot{\varepsilon}}{\partial \sigma} \tag{3}$$

where $k_B$ is the Boltzmann constant and $T$ is the temperature. With the Burgers vector **b** of 0.255 nm for a perfect dislocation of copper, $V^*$ of copper micropillar at RT between $\dot{\varepsilon}$ of 0.001 and 1 s$^{-1}$ was calculated as $34 \pm 2$ **b**$^3$ corresponding typically to dislocation-TB interaction [57–59]. At RT between $\dot{\varepsilon}$ of 1 and 100 s$^{-1}$, we hypothesize that the applied stress is high enough for the simultaneous movement of multiple dislocations, leading to enhanced interactions between them. These dislocation movements may lead to dislocation-TB interaction that occurs throughout the pillar [60,61]. The corresponding hardening modulus of micropillars tested at RT in Fig. S15 shows a constant increase with strain rate, which is attributed to the increased number of dislocation and TB interactions. As such, compared to the low $\dot{\varepsilon}$, the deformation is relatively homogeneous, but the dislocation glide can still occur occasionally and leave slip steps at the surface. This is also reflected in the apparent activation volume of $959 \pm 190$ **b**$^3$ calculated in this regime and it is quite close to the value for a typical forest dislocation obstructed movement. At CT between $\dot{\varepsilon}$ of 0.001 and 1 s$^{-1}$, mechanical twinning may occur at multiple grains (activation volume of $73 \pm 22$ **b**$^3$) due to the low temperature and high stresses involved, leading to homogeneous deformation without any slip steps at the surface. The hardening modulus of micropillars tested at CT shows relatively constant values regardless of the strain rate increase.

**Deformation behavior of microlattice based on micropillar properties**

Finally, the deformation behavior of the copper microlattices can be elucidated from the viewpoint of the deformation mechanism identified for copper micropillars. A difference in the relative strength of microlattices compressed at RT and CT can be attributed to the global mechanical response originating from the intrinsic deformation mechanisms at the component level, which is the strut. During the initial deformation, the struts at the outer surfaces are prone to deform first due to lower nodal connectivity. We hypothesize that, at CT, the plastic deformation is governed by mechanical twinning in multiple grains, leading to homogeneous deformation of the strut without introducing any mechanical instabilities as shown in Fig. 2(e).

On the other hand, at RT, the plastic deformation governed by dislocation-twin interactions and slips is expected (especially at $\dot{\varepsilon}$ <1 s$^{-1}$), leading to a relatively localized deformation of struts and slip steps at the surface. The localized deformation can act as a mechanical instability during deformation and further lead to buckling deformation of the strut, which is already confirmed by the stress overshoot and decrease after elastic deformation in the stress-strain curves (Fig. 2(c)). An upper elastic limit for the buckling of the strut can be given by the Euler buckling theory as follows:

$$\sigma_{cr} = \frac{\pi^2 E}{4}\left(\frac{R}{L}\right)^2 \quad (4)$$

where $\sigma_{cr}$ is critical buckling strength, $E$ is Young's modulus (110 GPa for copper), $R$ is the radius of the strut, and $L$ is the length of the strut. Eqn. (4) gives $\sigma_{cr}$ of 603 MPa which is higher than the yield strength of copper micropillars at any $\dot{\varepsilon}$ and temperature, suggesting that the buckling of the strut in the microlattice is not favorable during compression. However, it should be noted that an introduction of surface imperfections or local plasticity is known to lower $\sigma_{cr}$ [62–64]. Ranjbartoreh and Wang reported that topological defects decrease $\sigma_{cr}$ of carbon nanotube by 30 % [64]. From this perspective, a rough estimate of $\sigma_{cr}$ of a strut with a slip step at the surface is about 422 MPa, which is comparable to the yield strength of the copper micropillar at RT. A dynamic increase factor (DIF, $\sigma_{dynamic}/\sigma_{quasi-static}$) calculation indirectly supports the proposed deformation mechanism of microlattice (Fig. S16). On the one hand, the DIFs of the microlattice and the micropillar show a clear difference at RT, suggesting that the deformation behavior is different. In the micropillar, simple uniaxial compression leads to surface slip steps, while in the microlattice, elastic or elastoplastic buckling of the strut, which is facilitated by the surface slip steps, and subsequent stretch-dominated deformation is expected (Fig. 2(e)). On the other hand, the DIFs of the microlattice and the micropillar tested at CT show negligible differences. This is because, at CT, the compression leads to homogeneous deformation in the strut, leading to higher yield and flow strength of the microlattice compared to RT.

## 3. Conclusion

Mechanical responses of copper microlattices under various strain rates and temperatures were investigated. The microlattices were fabricated via the LEL process, with smooth surfaces and nodal connectivity close to the original design. They exhibited deformation behaviors dependent on temperature and strain rate during *in situ* micromechanical testing. A detailed study on the deformation mechanism of micropillars—having the same diameter and length as the struts of the microlattice revealed significant differences in the deformation mechanism based on temperature and strain rate. These distinct temperature-and rate-dependent deformation behaviors of the microlattice can be attributed to the collective mechanical response of the microlattice under compression, stemming from the intrinsic deformation mechanisms present at the component level, specifically the strut. This comprehensive understanding provides insights into the nuanced yet tunable responses and functionalities of such metallic microstructures under demanding operational conditions. This study represents the first step towards future bottom-up additive microfabrication of metal MEMS and microelectronics, as well as a new methodology for physical micrometallurgy explorations.

# Methods

## Printing of microarchitectures

Copper microarchitectures in this study were fabricated by LEL via localized force-controlled electrodeposition using the CERES system (Exaddon AG, Switzerland). An electrolyte containing copper ions flows through a 300-nm-diameter orifice in a silicon nitride AFM tip immersed in a standard three-cell electrochemical cell by applied pressure. The electrolyte contains 0.5 M $CuSO_4$ in 51 mM $H_2SO_4$ and 0.48 mM HCl, with brightener, leveler and surfactant added (Printing Solution Bright [Cu], Exaddon AG, Switzerland). Subsequently, the ions are deposited at desired locations on a Si/Ti (10 nm)/Cu (100 nm) substrate contacting the working electrode. The deposition is done in a voxel-by-voxel manner such that the AFM cantilever moves to the next voxel coordinate when it is deflected by the growing copper voxel, whereby the deflection is registered using standard optical beam deflection. The deflection threshold is in the order of 1 nm. The pressure and applied potential of LEL for microlattice printing were 20 mbar and -0.5 V, respectively. For micropillar printing, the applied potential was -0.5 V, and the pressure was carefully controlled in the range of 20 to 70 mbar to obtain a consistent diameter in the gauge section. To get smooth surfaces, the distance between each voxel was kept at 0.5 μm. Previous studies have already shown that the printed copper parts printed via LEL show dense microstructure and smooth surfaces [56]. Detailed information on LEL is given elsewhere [65].

## Microstructural characterization

The microstructure of copper microlattices and micropillars was characterized via SEM-TKD and STEM analysis. The cross-sections parallel to the deposition direction (along z-axis) were sampled using focused-ion-beam milling (Thermo Fisher Scios 2 dual beam FIB/SEM). TKD patterns were captured in a Zeiss Sigma SEM operated at 20 kV with a step size of 20 nm. A critical misorientation angle for grain boundaries was set as 15° to distinguish between low-angle grain boundary (LAGB) and high-angle grain boundary (HAGB). Twin boundaries were identified according to the rotation axes and angles, with a deviation tolerance of 5°. Scanning transmission electron microscopy (STEM) was conducted on a Titan Themis microscope operated at 300 kV. The convergence semi-angle of the aberration-corrected probe is 23.8 mrad to reach a probe size of 0.1 nm. High-angle annular dark field (HAADF) STEM images were taken using a detector range of 73~200 mrad.

**Nano-CT**

The nano-CT experiments were performed utilizing a lab-based ZEISS Xradia 810 Ultra X-ray microscope equipped with a 5.4 keV (Cr Kα) rotating anode source. Due to the relatively strong X-ray attenuation of copper, the samples were investigated using absorption contrast in the large-field-of-view (LFOV) mode of the instrument, offering a resolution of about 150 nm and a field of view of 64 µm x 64 µm. Using the ZEISS Xradia 810 Ultra's native software ZEISS Scout&Scan, 761 projections were acquired with an exposure time of 300 s each. Three-dimensional reconstruction of the acquired tilt series was performed by application of a simultaneous iterative reconstruction technique (SIRT) implementation (100 iterations) based on the ASTRA toolbox [66,67]. We assume X-ray attenuation according to Beer-Lambert law, therefore we preprocessed the input projection data before SIRT reconstruction (so-called cupping artifact correction) using projection data $p = -ln(I/I_0)$, where $I$ is the measured X-ray intensity and $I_0$ is the unattenuated incident X-ray intensity. However, due to the very high absorption of the dense copper, some non-linear contrast artifacts remain in the 3D reconstruction. Remaining artifacts may further be reduced by implementing more advanced reconstruction algorithms as recently demonstrated by Kreuz *et al* [68]. The reconstructed volumes were visualized and analyzed with the software arivis Vision4D 4.1.0.

*In situ* **micromechanical test**

The compression experiments at room temperature and low temperature were performed *in situ* using a micromechanical testing system (Alemnis AG, Switzerland). To perform the *in situ* tests, the indenter was installed in a Zeiss Gemini 500 SEM (Zeiss, Germany) and a JEOL JSM-6490 (JEOL Instruments, Japan). Experiments were performed in both setups in displacement control using a flat punch diamond indenter of 150 µm diameter for the microlattices and 5 µm for the micropillars (Synton-MDP AG, Switzerland). For quasi-static testing, the testing setup uses the piezoelectric actuator that enables intrinsic displacement control and a strain gauge load cell that limits the actuation speeds to 10 µm/s due to its susceptibility to ringing. To enable mechanical testing beyond speeds of 10 µm/s, a piezo-based load cell was employed instead of the strain gauge load cell, which extends the maximum achievable testing speed to 10 mm/s (limited by the voltage amplifier's ability to supply high voltages at high speeds to the piezoactuator). Given the short time scales of the experiments, data acquisition systems with a high acquisition frequency upto ~1 MHz were utilized to record the load and displacement data. Furthermore, to mitigate the potential impact of resonance from low stiffness springs in the testing system on the load data, load measurements were taken both with and without the sample under identical high speed actuation conditions. Subsequently, the load corresponding only to the sample, devoid of any resonance effects, was derived by subtracting the load recorded with the sample from the load recorded without the sample.

The micromechanical testing setup for cryogenic temperatures consists of a cold finger that is cooled by liquid nitrogen, which is pumped into the system from an outside dewar [28]. The cold finger is linked to the tip and the sample by copper braids and separated from the indenter frame by ceramic shafts. The tip, sample, and frame have individual resistive heaters and thermocouples in a closed feedback loop that enable precise control of the temperature in the system. To minimize drift, we maintained the frame at a constant temperature using the aforementioned heaters and temperature feedback. Once the system was cooled down, we switched on the frame heaters and let them stabilize until the change in frame temperature was less than $\pm 0.01°C$ in 10 minutes. Before performing actual compression on the microlattices and micropillars, we matched the temperature between the indenter and the sample to eliminate drift [69]. For this, we conducted several flat-punch indentations under load control, where the indenter or sample temperature was changed with respect to the other by $\pm 5°C$. We then measured the drift in a hold segment during the unload. We chose the pair of temperatures that provided a displacement drift lower than 100 pm/s.

**Finite element simulations**

Finite element simulations of the mechanical behavior of the copper microlattice were performed using Abaqus/Standard (Dassault Systèmes, France). Two configurations were considered: an idealized model of the complete lattice structure including a solid base and top as well as a configuration where the solid top layer was removed. The model was meshed using 1.5 million quadratic tetrahedral elements (C3D10). Nodes were fixed in all directions at the bottom of the base layer. Uniaxial compression was simulated by kinematically coupling all nodes of the top surface to a reference node on which displacement boundary conditions were applied. The material behavior of copper was modeled using an isotropic elastic ideal plastic material model based on a von Mises criterion. The elastic modulus, Poisson's ratio, and yield strength of copper were chosen as 120 GPa, 0.3, and 0.35 GPa, respectively. The reaction force was extracted from the reference node and converted to engineering stress by dividing it by the initial cross-sectional area of the lattice structure. Engineering strain was determined by dividing the relative reference node displacement by the initial height of the lattice structure.

**Acknowledgements**

S-G.K. was supported by the National Research Foundation of Korea (NRF, No. NRF-2020R1A6A3A03039038) and the research grant of the Gyeongsang National University in 2023. R.R. would like to acknowledge funding from the European Union (ERC Starting Grant, AMMicro, 101078619). Views and opinions expressed are however those of the author(s) only and do not necessarily reflect those of the European Union or the European Research Council. Neither the European Union nor the granting authority can be held responsible for them. S.K. was funded by the French National Research Agency (ANR) under the project No. "ANR-22-CE08-0012-01" (INSTINCT). A.G., J. W., B.A.Z and E.S., acknowledge financial support by the German Research Foundation (DFG) within the frameworks of the Collaborative Research Centre 1411 (Project-ID 416229255), the Collaborative Research Centre 1452 (Project-ID 431791331), and the project SP648/8 (Project-ID 316992193).

# References

1. Kumar Katiyar, N., Goel, G., Hawi, S. & Goel, S. Nature-inspired materials: Emerging trends and prospects. *NPG Asia Mater* **13:56**, (2021).

2. Chen, S., Chen, J., Zhang, X., Li, Z. Y. & Li, J. Kirigami/origami: unfolding the new regime of advanced 3D microfabrication/nanofabrication with "folding". *Light: Science and Applications* vol. 9 Preprint at https://doi.org/10.1038/s41377-020-0309-9 (2020).

3. Neville, R. M., Scarpa, F. & Pirrera, A. Shape morphing Kirigami mechanical metamaterials. *Sci Rep* **6:1**, 1–12 (2016).

4. Wu, Y. *et al.* Additively manufactured materials and structures: A state-of-the-art review on their mechanical characteristics and energy absorption. *Int J Mech Sci* **246**, 108102 (2023).

5. Riva, L., Ginestra, P. S. & Ceretti, E. Mechanical characterization and properties of laser-based powder bed-fused lattice structures: a review. *The International Journal of Advanced Manufacturing Technology* **113**, 649–671 (2021).

6. Bici, M. *et al.* Development of a multifunctional panel for aerospace use through SLM additive manufacturing. *Procedia CIRP* **67**, 215–220 (2018).

7. Zhou, J., Shrotriya, P. & Soboyejo, W. O. On the deformation of aluminum lattice block structures: from struts to structures. *Mechanics of Materials* **36**, 723–737 (2004).

8. Burton, H. E. *et al.* The design of additively manufactured lattices to increase the functionality of medical implants. *Materials Science and Engineering: C* **94**, 901–908 (2019).

9. Alabort, E., Barba, D. & Reed, R. C. Design of metallic bone by additive manufacturing. *Scr Mater* **164**, 110–114 (2019).

10. Feng, J., Liu, B., Lin, Z. & Fu, J. Isotropic octet-truss lattice structure design and anisotropy control strategies for implant application. *Mater Des* **203**, 109595 (2021).

11. Greer, J. R. & De Hosson, J. T. M. Plasticity in small-sized metallic systems: Intrinsic versus extrinsic size effect. *Prog Mater Sci* **56**, 654–724 (2011).

12. Rusinek, A., Rodríguez-Martínez, J. A. & Arias, A. A thermo-viscoplastic constitutive model for FCC metals with application to OFHC copper. *Int J Mech Sci* **52**, 120–135 (2010).

13. Kang, S. G. *et al.* Green laser powder bed fusion based fabrication and rate-dependent mechanical properties of copper lattices. *Mater Des* **231**, 112023 (2023).

14. Otto, F. *et al.* The influences of temperature and microstructure on the tensile properties of a CoCrFeMnNi high-entropy alloy. *Acta Mater* **61**, 5743–5755 (2013).

15. Fu, W. *et al.* Cryogenic mechanical behaviors of CrMnFeCoNi high-entropy alloy. *Materials Science and Engineering: A* **789**, 139579 (2020).

16. Ni, L. & De Boer, M. P. Sacrificial Materials and Release Etchants for Metal MEMS That Reduce or Eliminate Hydrogen-Induced Residual Stress Change. *Journal of Microelectromechanical Systems* **30**, 426–432 (2021).

17. Ozóg, P., Rutkowski, P., Kata, D. & Graule, T. Ultraviolet lithography-based ceramic manufacturing (UV-LCM) of the aluminum nitride (AlN)-based photocurable Dispersions. *Materials* **13**, (2020).

18. Ahn, B. Y. *et al.* Omnidirectional Printing of Flexible, Stretchable, and Spanning Silver Microelectrodes. *Science (1979)* **323**, 1590–1593 (2009).

19. Makrygianni, M., Kalpyris, I., Boutopoulos, C. & Zergioti, I. Laser induced forward transfer of Ag nanoparticles ink deposition and characterization. *Appl Surf Sci* **297**, 40–44 (2014).

20. Ishikawa, A., Tanaka, T. & Kawata, S. Improvement in the reduction of silver ions in aqueous solution using two-photon sensitive dye . *Appl. Phys. Lett* **89**, 113102 (2006).

21. Menétrey, M. *et al.* Targeted Additive Micromodulation of Grain Size in Nanocrystalline Copper Nanostructures by Electrohydrodynamic Redox 3D Printing. *Small* **18**, (2022).

22. Zhang, W. *et al.* Suppressed Size Effect in Nanopillars with Hierarchical Microstructures Enabled by Nanoscale Additive Manufacturing. *Nano Lett* **23**, 8162–8170 (2023).

23. Saccone, M. A., Gallivan, R. A., Narita, K., Yee, D. W. & Greer, J. R. Additive manufacturing of micro-architected metals via hydrogel infusion. *Nature* **612**, 685–690 (2022).

24. Oran, D. *et al.* 3D PRINTING 3D nanofabrication by volumetric deposition and controlled shrinkage of patterned scaffolds. *Science (1979)* **362**, 1281–1285 (2018).

25. Bernardeschi, I., Ilyas, M. & Beccai, L. A Review on Active 3D Microstructures via Direct Laser Lithography. *Advanced Intelligent Systems* **3**, 2100051 (2021).

26. Wendy Gu, X. & Greer, J. R. Ultra-strong architected Cu meso-lattices. *Extreme Mech Lett* **2**, 7–14 (2015).

27. Khaderi, S. N. *et al.* The indentation response of Nickel nano double gyroid lattices. *Extreme Mech Lett* **10**, 15–23 (2017).

28. Schwiedrzik, J. *et al.* Dynamic cryo-mechanical properties of additively manufactured nanocrystalline nickel 3D microarchitectures. *Mater Des* **220**, 110836 (2022).

29. Lin, C.-C. & Hu, C.-C. The Ultrahigh-Rate Growth of Nanotwinned Copper Induced by Thiol Organic Additives. *J Electrochem Soc* **167**, 082505 (2020).

30. Lai, C. Q. & Daraio, C. Highly porous microlattices as ultrathin and efficient impact absorbers. *Int J Impact Eng* **120**, 138–149 (2018).

31. Kai, Y. *et al.* Dynamic diagnosis of metamaterials through laser-induced vibrational signatures. *Nature* **623**, 514–521 (2023).

32. Ramachandramoorthy, R. *et al.* High strain rate in situ micropillar compression of a Zr-based metallic glass. *J Mater Res* **36**, 2325–2336 (2021).

33. Jost, E. W., Moore, D. G. & Saldana, C. Evolution of global and local deformation in additively manufactured octet truss lattice structures. *Additive Manufacturing Letters* **1**, 100010 (2021).

34. Tancogne-Dejean, T., Spierings, A. B. & Mohr, D. Additively-manufactured metallic micro-lattice materials for high specific energy absorption under static and dynamic loading. *Acta Mater* **116**, 14–28 (2016).

35. He, Z. Z., Wang, F. C., Zhu, Y. B., Wu, H. A. & Park, H. S. Mechanical properties of copper octet-truss nanolattices. *J Mech Phys Solids* **101**, 133–149 (2017).

36. Daum, B. *et al.* Elastoplastic buckling as source of misinterpretation of micropillar tests. *Acta Mater* **61**, 4996–5007 (2013).

37. Qin, Z., Zhu, J., Li, W., Xia, Y. & Zhou, Q. System ringing in impact test triggered by upper-and-lower yield points of materials. *Int J Impact Eng* **108**, 295–302 (2017).

38. Schaedler, T. A. *et al.* Ultralight Metallic Microlattices. *Science (1979)* **334**, 962–965 (2011).

39. Cheng, H. *et al.* Mechanical metamaterials made of freestanding quasi-BCC nanolattices of gold and copper with ultra-high energy absorption capacity. *Nat Commun* **14**, (2023).

40. Bonatti, C. & Mohr, D. Large deformation response of additively-manufactured FCC metamaterials: From octet truss lattices towards continuous shell mesostructures. *Int J Plast* **92**, 122–147 (2017).

41. Mieszala, M. *et al.* Micromechanics of Amorphous Metal/Polymer Hybrid Structures with 3D Cellular Architectures: Size Effects, Buckling Behavior, and Energy Absorption Capability. *Small* **13**, (2017).

42. Zhang, X. *et al.* Three-Dimensional High-Entropy Alloy-Polymer Composite Nanolattices That Overcome the Strength-Recoverability Trade-off. *Nano Lett* **18**, 4247–4256 (2018).

43. Bauer, J., Hengsbach, S., Tesari, I., Schwaiger, R. & Kraft, O. High-strength cellular ceramic composites with 3D microarchitecture. *Proc Natl Acad Sci U S A* **111**, 2453–2458 (2014).

44. Ashby, M. F. The properties of foams and lattices. *Philosophical Transactions of the Royal Society A: Mathematical, Physical and Engineering Sciences* **364**, 15–30 (2006).

45. Bahrami Babamiri, B., Askari, H. & Hazeli, K. Deformation mechanisms and post-yielding behavior of additively manufactured lattice structures. *Mater Des* **188**, 108443 (2020).

46. Doutre, P.-T., Grandvallet, C., Gobet, L., Vignat, F. & Dendievel, R. Influence of fillets onto mechanical properties of octet-truss lattice structures. (2023).

47. Malyar, N. V, Grabowski, B., Dehm, G. & Kirchlechner, C. Dislocation slip transmission through a coherent S3{111} copper twin boundary: Strain rate sensitivity, activation volume and strength distribution function. *Acta Mater* **161**, 412–419 (2018).

48. Zhu, Y. T. *et al.* Dislocation–twin interactions in nanocrystalline fcc metals. *Acta Mater* **59**, 812–821 (2011).

49. Chassagne, M., Legros, M. & Rodney, D. Atomic-scale simulation of screw dislocation/coherent twin boundary interaction in Al, Au, Cu and Ni. *Acta Mater* **59**, 1456–1463 (2011).

50. Malyar, N. V., Micha, J. S., Dehm, G. & Kirchlechner, C. Dislocation-twin boundary interaction in small scale Cu bi-crystals loaded in different crystallographic directions. *Acta Mater* **129**, 91–97 (2017).

51. Wang, J. *et al.* Detwinning mechanisms for growth twins in face-centered cubic metals. *Acta Mater* **58**, 2262–2270 (2010).

52. Cao, Y. *et al.* Grain boundary formation by remnant dislocations from the de-twinning of thin nano-twins. *Scr Mater* **100**, 98–101 (2015).

53. Blewitt, T. H., Coltman, ; R R, Redman, ; J K, Coltman, R. R. & Redman, J. K. Low-Temperature Deformation of Copper Single Crystals. *J Appl Phys* **28**, 651–660 (1957).

54. Sarma, V. S. *et al.* Role of stacking fault energy in strengthening due to cryo-deformation of FCC metals. *Materials Science and Engineering: A* **527**, 7624–7630 (2010).

55. Meyers, M. A., Vöhringer, O. & Lubarda, V. A. The onset of twinning in metals: a constitutive description. *Acta Mater* **49**, 4025–4039 (2001).

56. Ramachandramoorthy, R. *et al.* Anomalous high strain rate compressive behavior of additively manufactured copper micropillars. *Appl Mater Today* **27**, 101415 (2022).

57. Wei, Q., Cheng, S., Ramesh, K. T. & Ma, E. Effect of nanocrystalline and ultrafine grain sizes on the strain rate sensitivity and activation volume: fcc versus bcc metals. *Materials Science and Engineering: A* **381**, 71–79 (2004).

58. Wang, Y. M. & Ma, ; E. Temperature and strain rate effects on the strength and ductility of nanostructured copper. *Appl. Phys. Lett* **83**, 3165–3167 (2003).

59. Lu, L. *et al.* Nano-sized twins induce high rate sensitivity of flow stress in pure copper. *Acta Mater* **53**, 2169–2179 (2005).

60. Khantha, M., Vitek, V. & Pope, D. P. Strain-rate dependent mechanism of cooperative dislocation generation: application to the brittle–ductile transition. *Materials Science and Engineering: A* **319–321**, 484–489 (2001).

61. Jennings, A. T., Li, J. & Greer, J. R. Emergence of strain-rate sensitivity in Cu nanopillars: Transition from dislocation multiplication to dislocation nucleation. *Acta Mater* **59**, 5627–5637 (2011).

62. Mousavi Davoudi, S. A. & Naghipour, M. Presentation of Critical Buckling Load Correction Factor of AISC Code on L-Shaped Composite Columns by Numerical and Experimental Analysis. *International Journal of Steel Structures* **20**, 1682–1702 (2020).

63. Lee, A., Jiménez, F. L., Marthelot, J., Hutchinson, J. W. & Reis, P. M. The Geometric Role of Precisely Engineered Imperfections on the Critical Buckling Load of Spherical Elastic Shells. *Journal of Applied Mechanics, Transactions ASME* **83**, (2016).

64. Ranjbartoreh, A. R. & Wang, G. Effect of Topological Defects on Buckling Behavior of Single-walled Carbon Nanotube. *Nanoscale Res Lett* **6:28**, (2010).

65. Ercolano, G. *et al.* Additive manufacturing of sub-micron to sub-mm Metal structures with hollow AFM cantilevers. *Micromachines (Basel)* **11**, (2020).

66. van Aarle, W. *et al.* The ASTRA Toolbox: A platform for advanced algorithm development in electron tomography. *Ultramicroscopy* **157**, 35–47 (2015).

67. Van Aarle, W. *et al. Performance Improvements for Iterative Electron Tomography Reconstruction Using Graphics Processing Units {(GPUs)}. CIRP Annals-Manufacturing Technology* vol. 157 http://web.eecs.umich.edu/~fessler/code/.14."TheASTRAToolbox,"http://www.astra-toolbox.com. (2010).

68. Kreuz, S. *et al.* Improving reconstructions in nanotomography for homogeneous materials via mathematical optimization. (2023).

69. Wheeler, J. M., Armstrong, D. E. J., Heinz, W. & Schwaiger, R. High temperature nanoindentation: The state of the art and future challenges. *Curr Opin Solid State Mater Sci* **19**, 354–366 (2015).

**Figure 1.** (a) Copper microarchitectures fabricated by localized electrodeposition process. (b) Computer-aided design of microlattice. Cube with white dashed line indicates a unit-cell. (c) Discretized voxel locations used to print microlattice. Red box indicates magnified area below showing schematic of voxel diameter and overlap. (d) SEM image of as-fabricated copper microlattice. (e) Reconstructed microlattice and unit-cell by NanoCT. Cube with red dashed line indicates a unit-cell. (f) Strut diameter distribution in as-fabricated microlattice. (g and h) Inverse pole figure and grain boundary map overlayed with image quality map of representative strut and node unit cross-section in microlattice obtained using TKD.

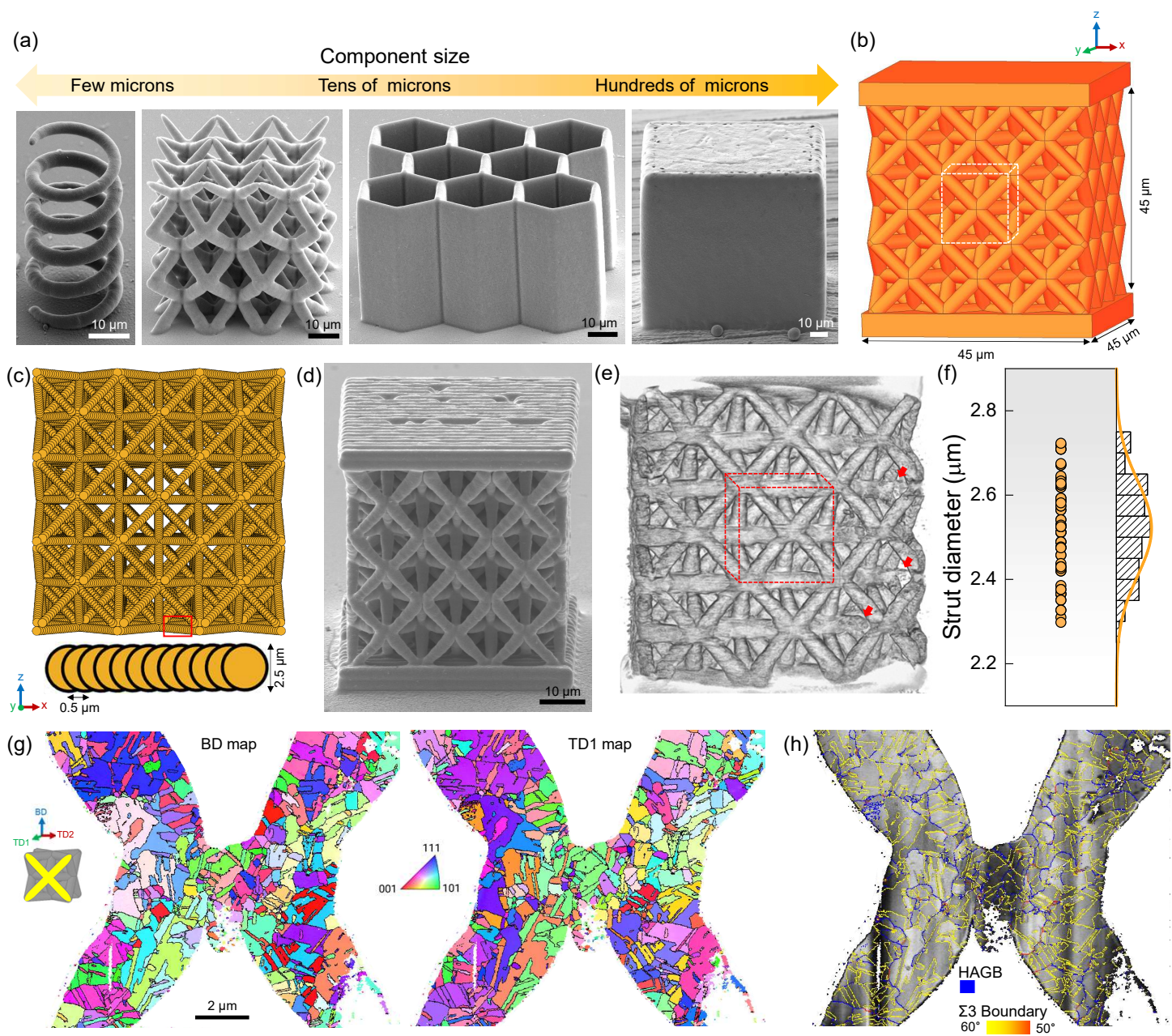

**Figure 2.** Compression testing of copper microlattices. (a and b) SEM images of microlattices during compression at RT and CT, respectively at a strain rate of 0.01 s$^{-1}$. (c) Engineering stress-strain curves of microlattices compressed at various temperatures and strain rates (Zoomed in inset: Representative stress-strain curves for RT and CT upto strain rates of 0.1/s). (d) Yield strength of microlattices as a function of strain rate at RT and CT. (e and f) SEM images of compressed microlattices at different strain rates with zoomed in insets. (g) Comparison of microlattice specific energy absorption between current study and previous literature.

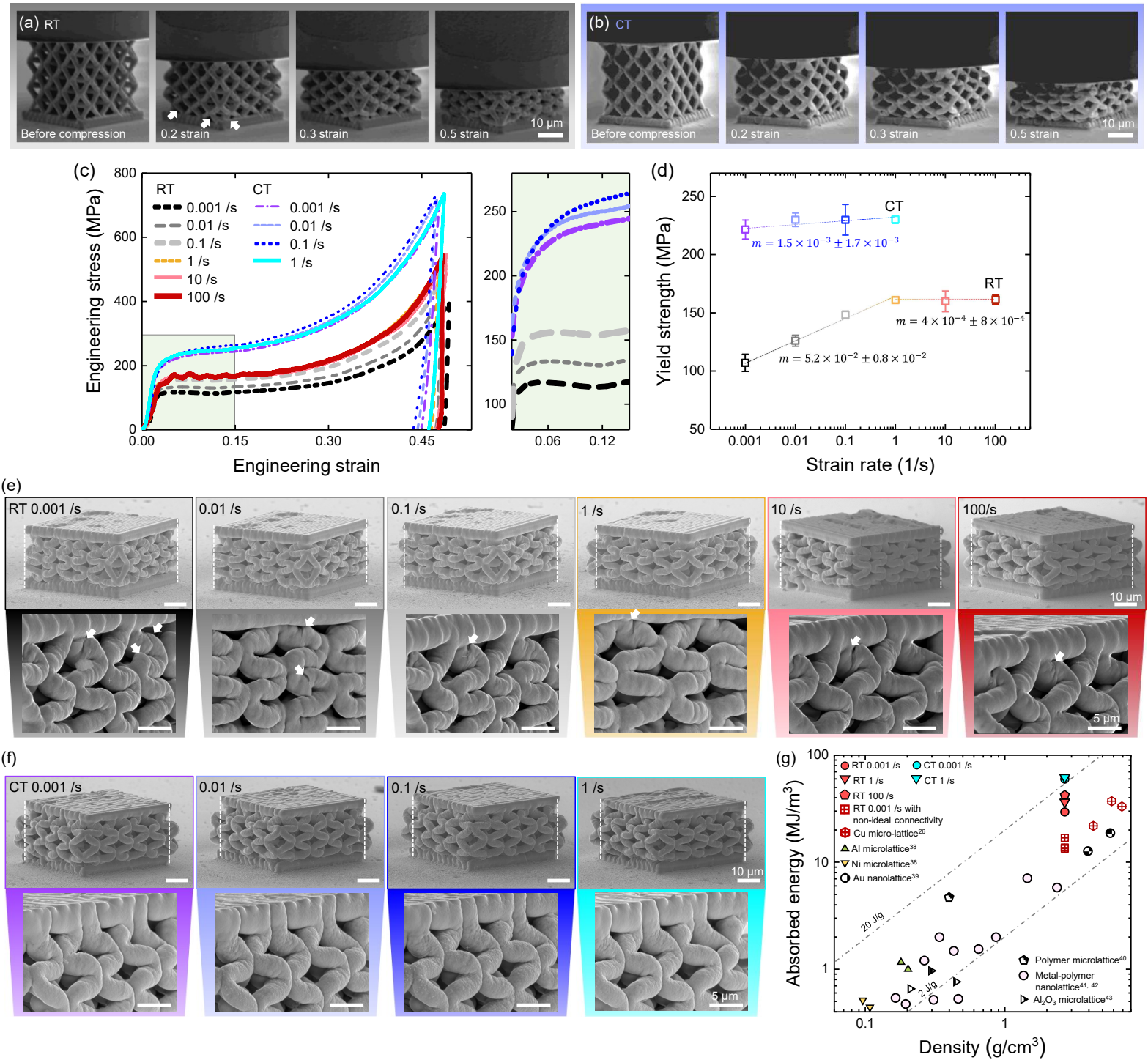

**Figure 3.** Copper micropillars: microstructural and mechanical characterization. (a) Computer-aided design of micropillar. (b) Discretized voxel locations used to print the micropillar. (c) SEM image of micropillars. White box indicate a region of interest for cross-sectional microstructure analysis (d) Inverse pole figure and grain boundary map of cross-section of micropillar along BD and TD1 obtained using TKD. (e) Engineering stress-strain curves of micropillars compressed at various temperatures and strain rates (Zoomed in inset: Stress-strain curves for CT upto strain rates of 1/s showing stress drops). (f) Yield strength of micropillars as a function of strain rate at RT and CT. (g) SEM images of compressed micropillars.

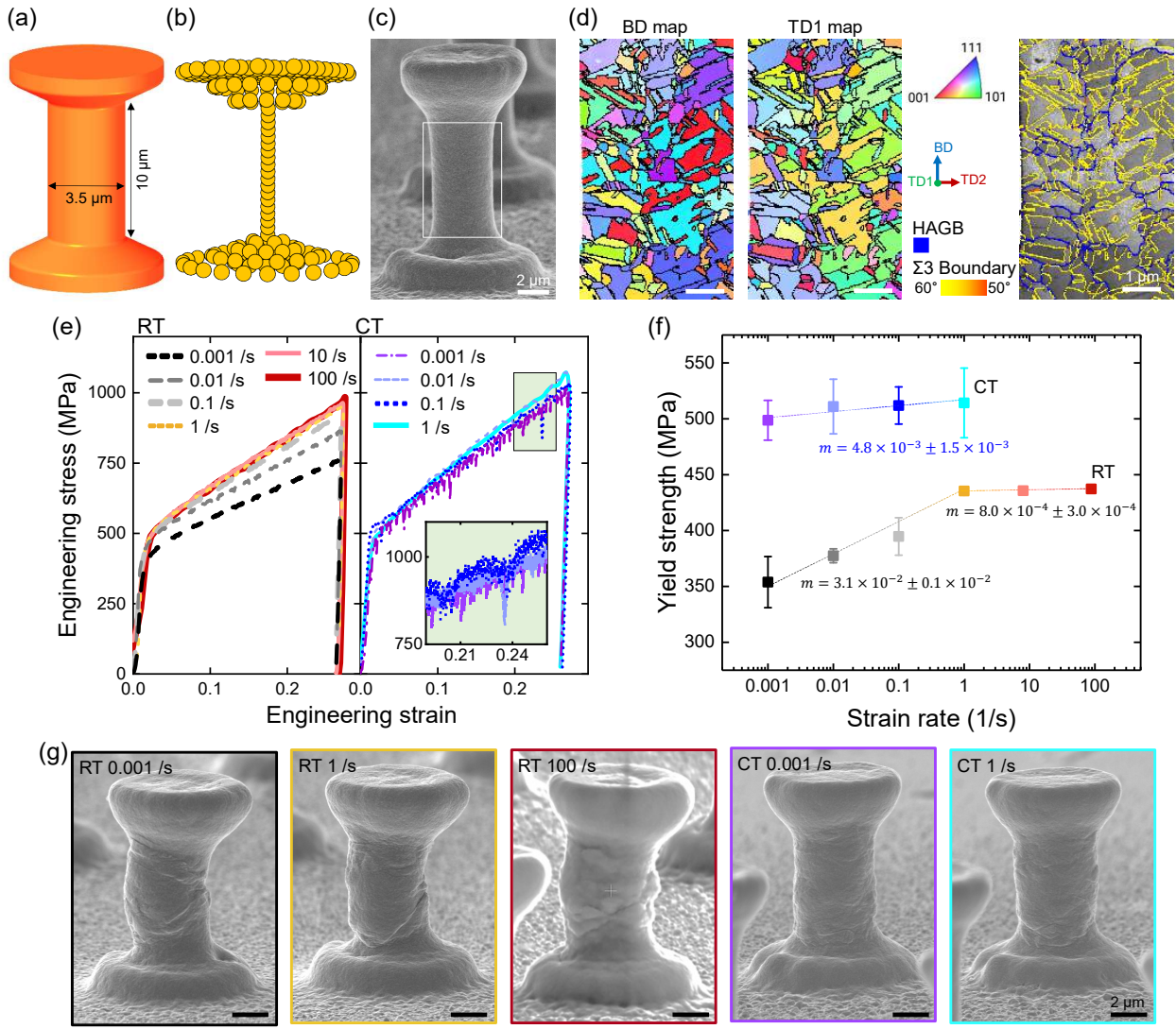

**Figure 4.** Deformation mechanism in copper micropillars. (a and b) Inverse pole figure and grain boundary map of cross-section of compressed micropillar. (c) HAADF-STEM image and FFT patterns obtained from atomically resolved areas with straight twin boundaries in micropillar compressed at CT with a $\dot{\varepsilon}$ of 0.001 /s. (d) Misorientation angle distribution of micropillar without compression (top), compressed at 0.001 /s (middle) and compressed at 1 /s (bottom) at RT and CT. (e) Stress drop magnitude with respect to drop time obtained from stress-strain curves of micropillars compressed at CT with a $\dot{\varepsilon}$ of 0.001, 0.01, and 0.1 /s. Inset is an zoomed-in engineering stress-time plot from micropillar compressed at CT with a $\dot{\varepsilon}$ of 0.01 /s. (f) Schematics of postulated deformation mechanisms in copper micropillar at different rates and temperatures.

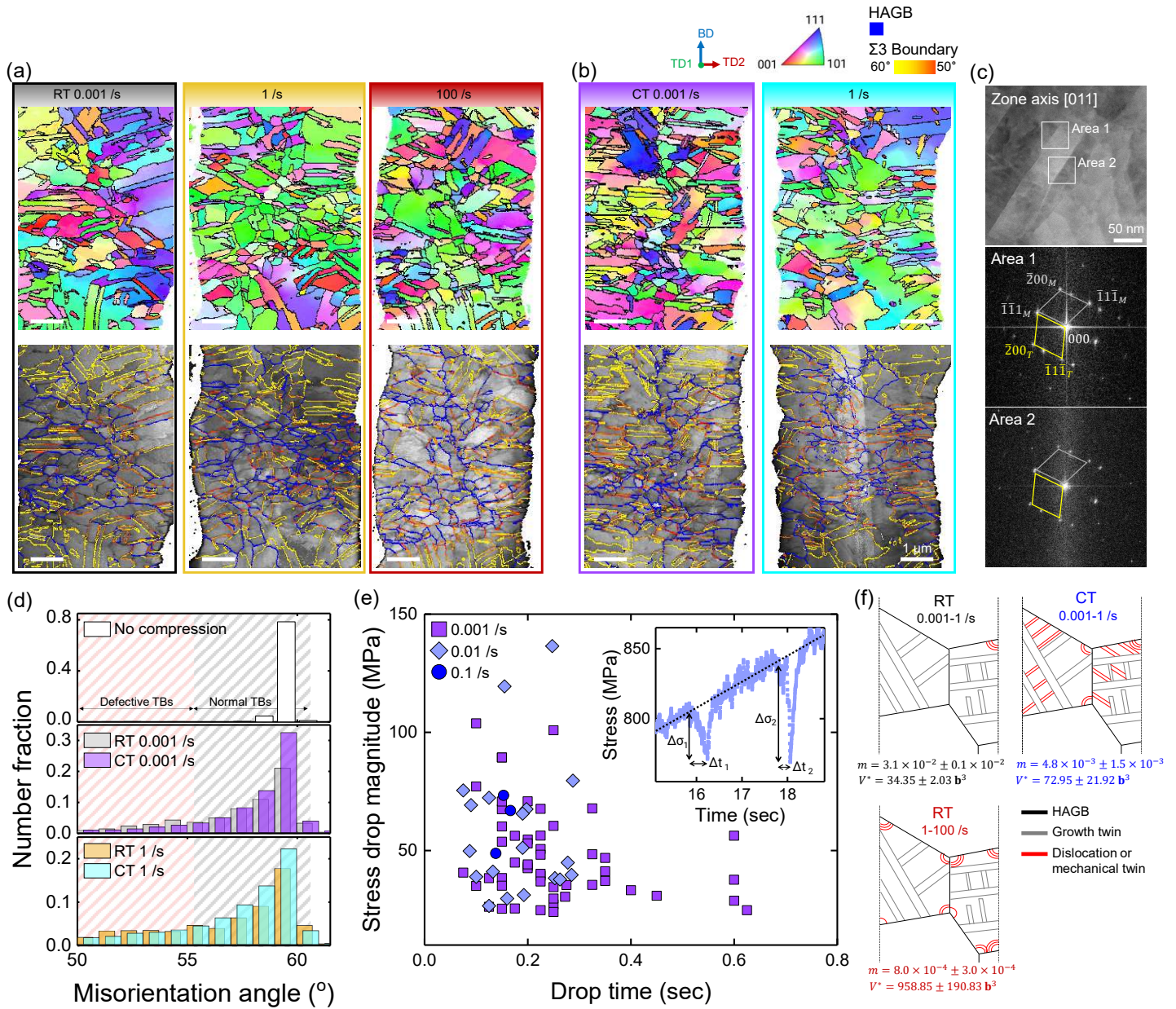

**Figure S1.** SEM images of Copper microlattices fabricated with different printing overpotentials at the same pressure of 20mbar.

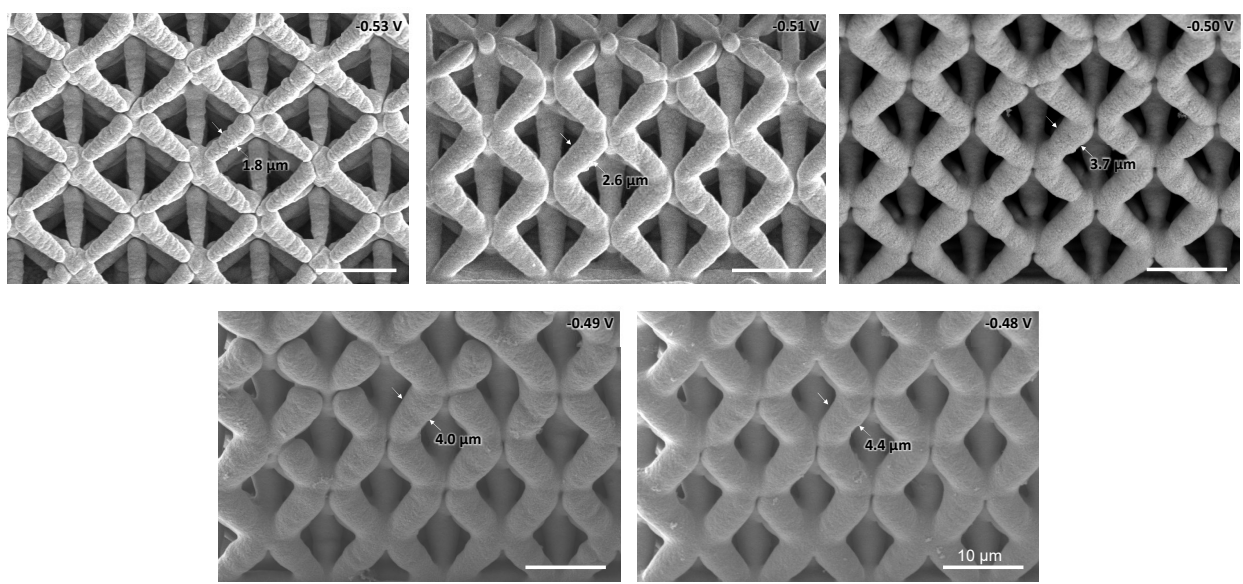

**Figure S2.** X-ray absorption contrast nano-CT images of a copper microlattice. (a) Front view (0°) projection image of the structure with solid matter bright and air/void spaces dark (inverted contrast). The interconnected structure and channels are well visible. Projection images of the sample at (b) 45° and (c) 22.5° rotation showing the channels and structure from other directions (see also Movie M1). (d, e) Horizontal and (f) vertical virtual cross-sectional slices through the 3D reconstruction (see also Movies M2 and M3) revealing the regular and integral strut morphology (bright contrast).

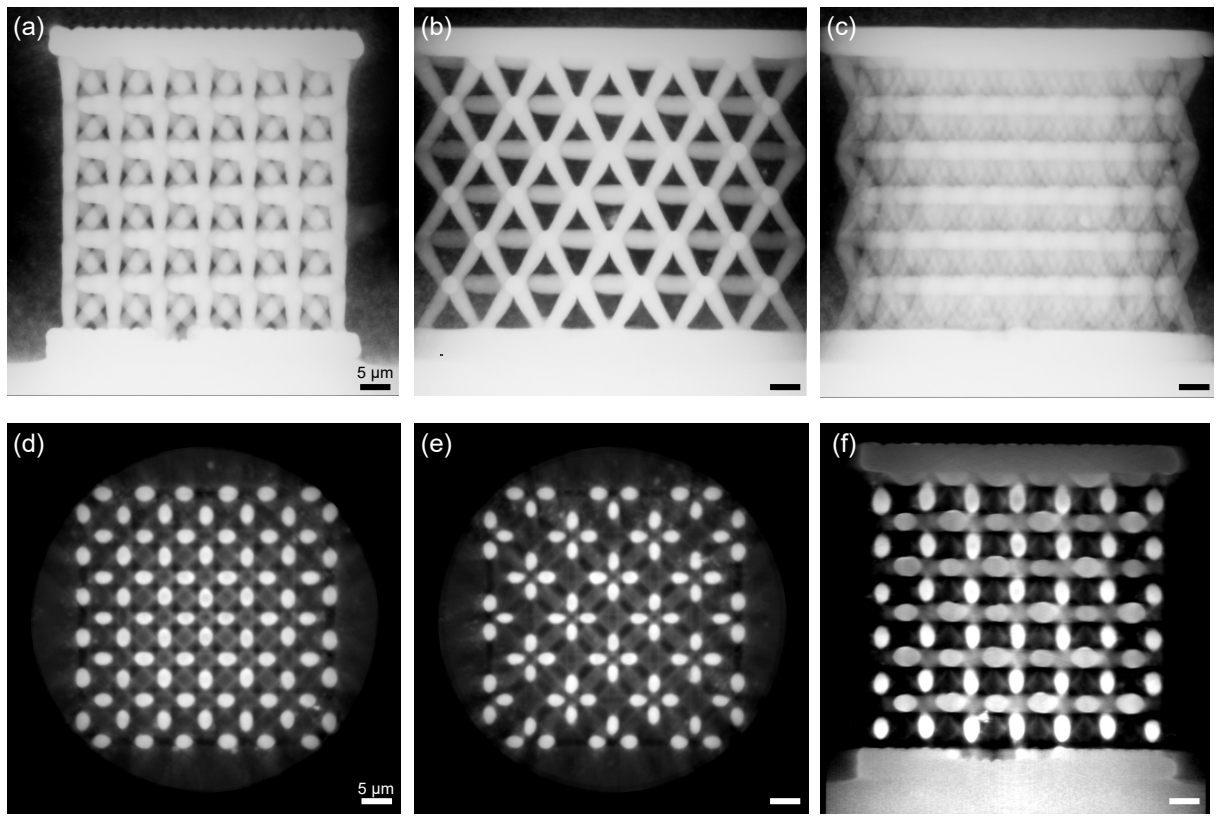

**Figure S3.** Image contrast map of the representative unit cross-section along the building direction of the microlattice (Same area presented in Fig. 1(g, h).

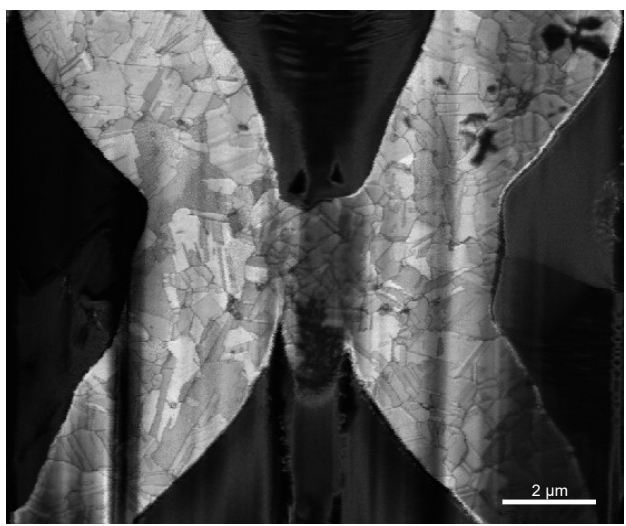

**Figure S4.** Flow stress (at 0.2 engineering strain) of microlattices at different temperatures and strain rate conditions.

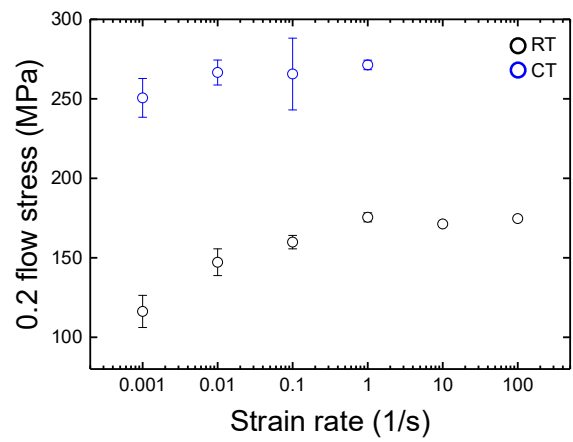

**Figure S5.** Stress-strain curves of microlattices. (a) At RT with a $\dot{\varepsilon}$ of 0.001, 0.01, 0.1, 1, 10, 100 s$^{-1}$. (b) At CT with a $\dot{\varepsilon}$ of 0.001, 0.01, 0.1, 1 s$^{-1}$.

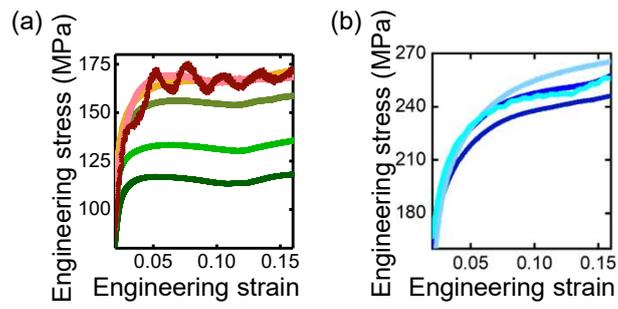

**Figure S6.** Degree of bulging out of compressed microlattices.

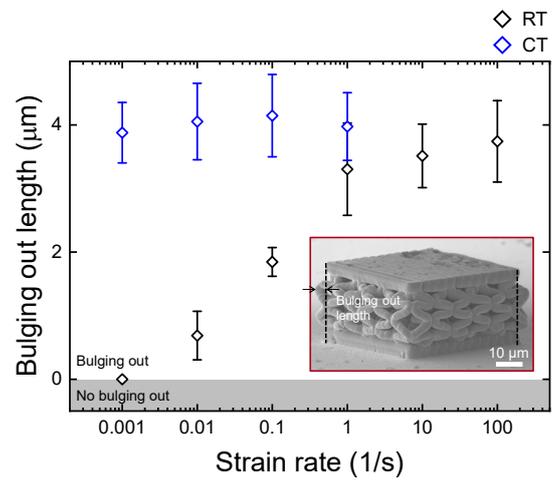

**Figure S7.** Deformation behavior of copper microlattices printed intentionally with poor nodal connectivity.

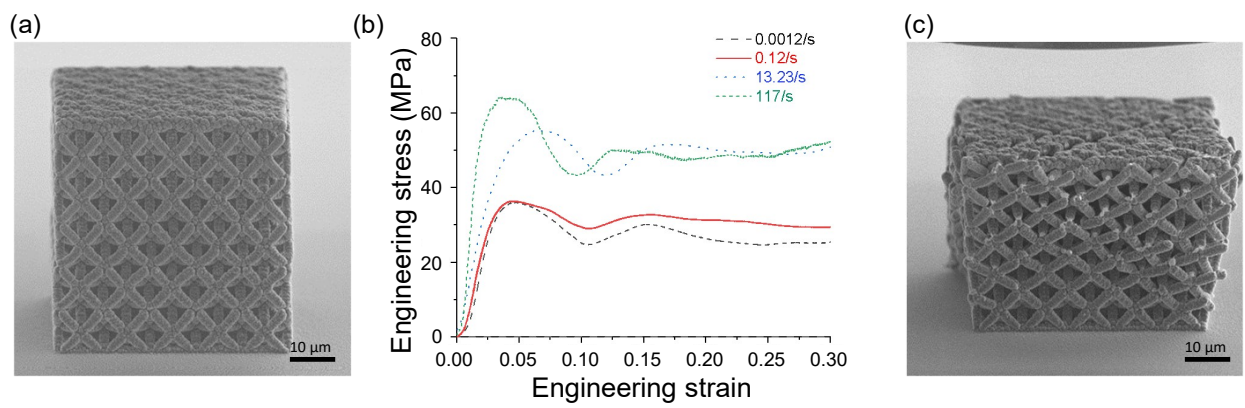

**Figure S8.** Finite element analysis of (a) Calculated compressive load-displacement curves of copper microlattices with and without top plate. (b) Plastic strain distribution in microlattices.

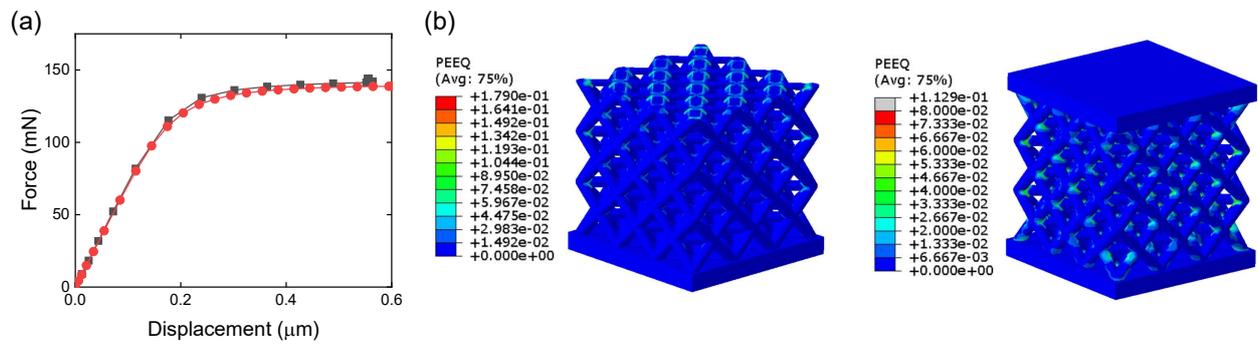

**Figure S9.** Strain rate map of microlattice cross-section calculated by FE simulation.

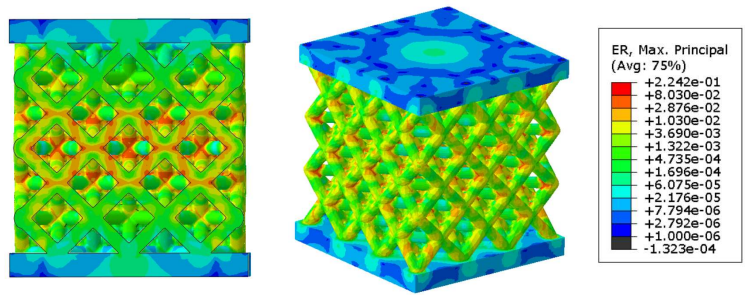

**Figure S10.** Misorientation angle distribution of microlattice and micropillar.

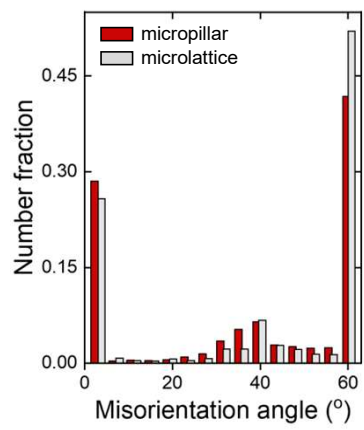

**Figure S11.** Flow stress (at 0.2 engineering strain) of micropillars at different temperatures and strain rate conditions.

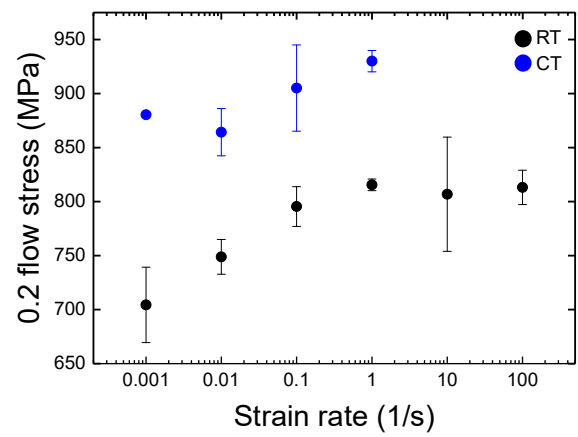

**Figure S12.** SEM images of compressed micropillars at different strain rates and temperatures.

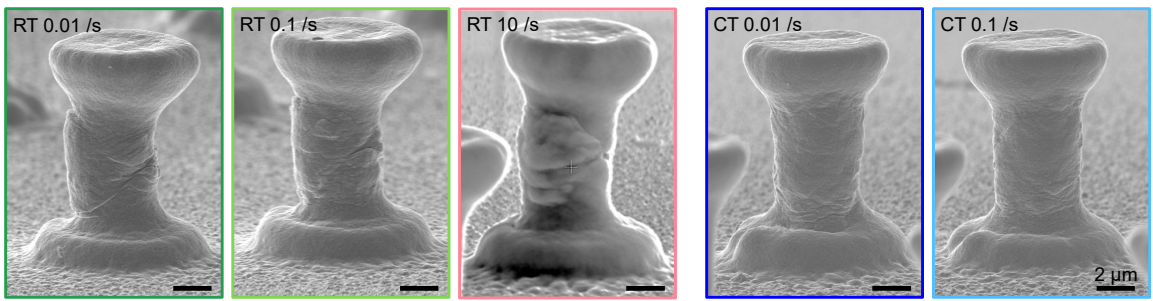

**Figure S13.** High resolution STEM image from copper micropillar compressed at CT with a $\dot{\varepsilon}$ of 0.001/s used for FFT pattern analysis.

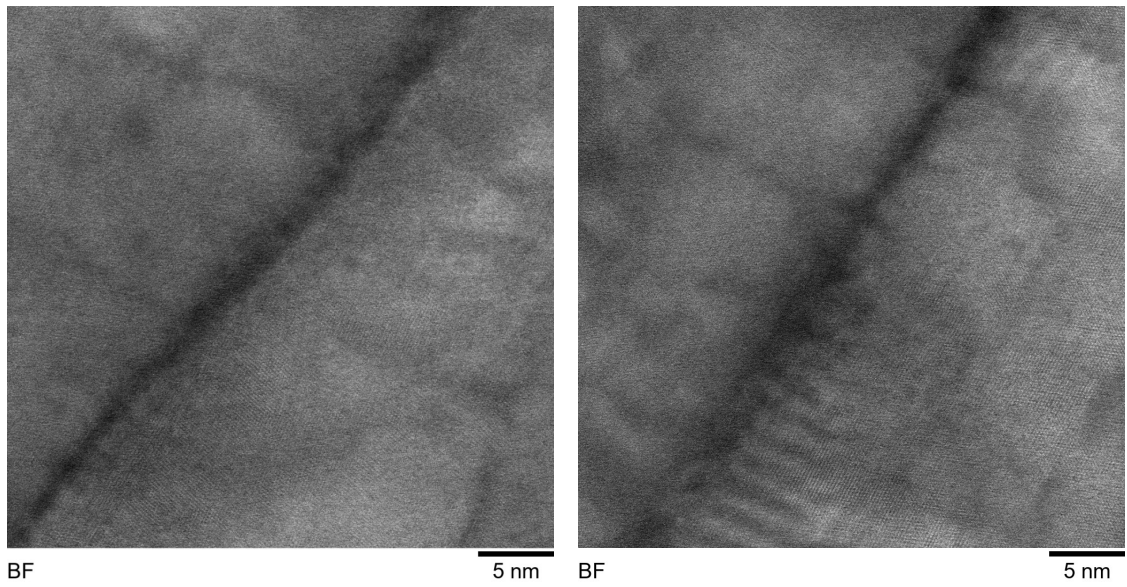

**Figure S14.** Grain boundary maps and corresponding pole figure maps of micropillar compressed at CT with a $\dot{\varepsilon}$ of 0.001/s.

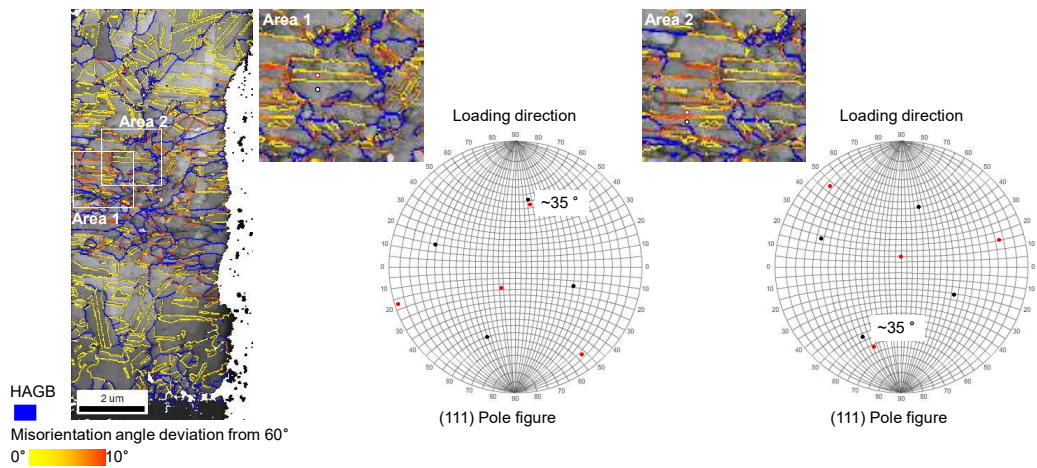

**Figure S15.** Hardening modulus of copper micropillars at cryogenic temperature (blue circles) and room temperature (black circles)

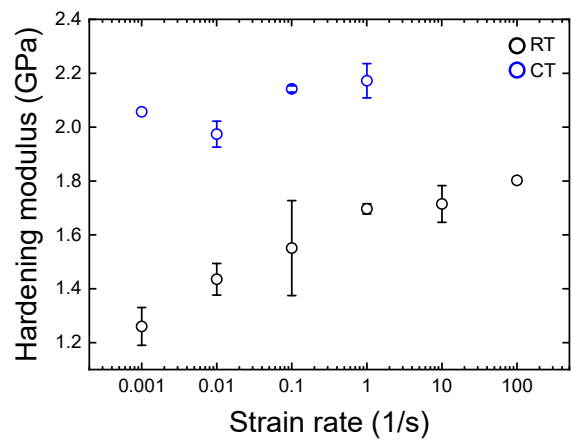

**Figure S16.** Dynamic increase factor of copper microlattices (empty squares) and copper micropillars (solid squares).

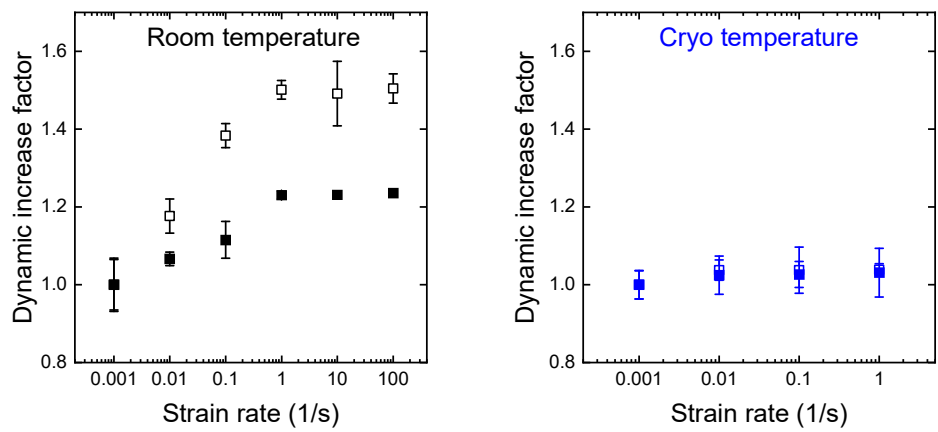